\documentclass[prb,letter,twocolumn,
showpacs,lengthcheck,superscriptaddress]{revtex4}
\usepackage{graphicx}
\usepackage{amsmath,amssymb}
\usepackage{bm}
\usepackage{ifthen}
\usepackage{dcolumn}

\begin{document}
\title{ Electronic structures of (In,Ga)As/GaAs quantum dot molecules
made of dots with dissimilar sizes}
\author{Lixin He}
\affiliation{Key Laboratory of Quantum Information, University of Science 
and Technology of China, Hefei, Anhui 230026, Peoples Republic of China}
\author{Alex Zunger}
\affiliation{National Renewable Energy Laboratory, Golden, Colorado 80401, USA}

\begin{abstract}

Using single-particle pseudopotential and many-particle configuration
interaction methods, 
we compare various physical quantities of (In,Ga)As/GaAs 
quantum dot molecules (QDMs) made of dissimilar dots 
(heteropolar QDMs) 
with QDMs made of identical dots (homopolar QDMs).
The calculations show that the electronic structures of 
hetero-QDMs and homo-QDMs differ significantly at large inter-dot distance.
In particular (i) Unlike those of homo-QDMs, the  single-particle molecular orbitals 
of hetero-QDMs convert to dot localized orbitals at large inter-dot distance. 
(ii) Consequently, in a hetero-QMD the 
bonding-antibonding splitting of molecular orbitals at
large inter-dot distance is
significantly larger than the electron  hopping energy  whereas
for homo-QDM, the bonding-antibonding splitting is very similar to the 
hopping energy.
(iii) The asymmetry of the QDM increases significantly the double
occupation for the two-electron ground states, and therefore 
lowers the degree of entanglement of the two electrons. 

\end{abstract}
\pacs{73.22.GK, 03.67.Mn, 85.35.-p}

\date{\today}
\maketitle

\section{Introduction}

Vertically coupled quantum dots\cite{xie95,solomon96} obtained via epitaxial
growth provide a potential scheme for scalable nano-structures for
quantum computing. In this scheme, two coupled quantum dots are
used as a basic logic gate, via the entanglement of 
one exciton\cite{bayer01} or two electronic spins.\cite{loss98} 
This proposal for gate operations, requires knowledge of the detailed physical 
properties of the ``quantum gate'' made of two quantum dots.
Significant progress has been recently made \cite{petta04,johnson05} 
using quantum dot molecules 
made of very large ($\sim$ 500 - 1000 \AA) 
electrostatically confined dots. 
The limit of large quantum confinement,
however, requires working with (200 $\times$ 30 \AA) 
self-assembled QDMs. 
So far, most experiments on self-assembled QDM are optical,\cite{bayer01} 
and most theories are based on
continuum models, such as effective mass approximations.\cite{bayer01}
These simple models ignore or drastically simplify important 
real material properties such as
strain, atomistic symmetries and crystal structural effects, band coupling
etc. Recent studies \cite{bester04b}
show that simplification of such important effects may lead to
qualitative changes in fundamental physics of the QDMs.

Previously, we have studied homopolar QDMs made of two identical quantum dots,
using single-particle pseudopotential method and
many-particle configuration interaction method.  \cite{he05b, he05c} 
We have studied electron localization, double occupation rate
and two-electron entanglement using a new 
formula for measuring the degree of entanglement formula for
two {\it indistinguishable} fermions. 
We found that even geometrically identical dots in the QDMs 
lead to electronic asymmetry due to the strain effects. 
However, experimentally it is hard to control the shape, size and
compositions of individual dots within the QDMs,
so in practice, the QDMs are never made of identical
dots. Actually, the top dots are tend to be larger 
than the bottom dots due to the strain effects.\cite{xie95,solomon96}
Indeed, the measured difference in exciton energy due to dot-size difference
is about 4 meV \cite{bodefeld99} for two vertically coupled dots that are 20 nm apart.
Sometimes, the two dots are intentionally grown different so that they can be
addressed separately. \cite{stinaff06} 
To provide quantitative comparison to experiments, considering the effects that asymmetry
of quantum dots within the molecule, we studied the QDMs made
of (In,Ga)As/GaAs quantum dots of different sizes (hetero-polar QDM).

In this paper, we study systematically the electronic properties of 
hetero-QDMs, including their single-particle molecular orbitals, 
many-particle states, double occupation and entanglement of two-electrons,
and compare them to those of homo-QDMs.
We found that while at {\it short} inter-dot distance, the electronic
properties of hetero-QDM and homo-QDM are similar, 
they differ significantly at {\it large} inter-dot distance.
This difference may have substantial impact in implementation of
quantum gates.    
      
\begin{figure}
\includegraphics[width=2.6in,angle=0]{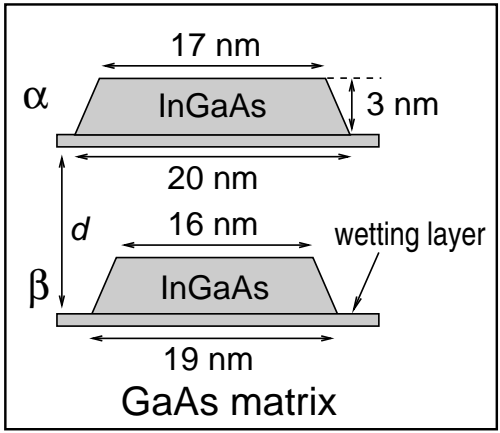}
\caption{The geometry used in this work for quantum dot molecules made of dissimilar dots.
We denote the (isolated) top dot ``$\alpha$'' and the (isolated) bottom dot
``$\beta$''. Each dot has the shape of a truncated cone. 
The inter-dot distance is measured from wetting layer to wetting layer.
}
\label{fig:geom}
\end{figure}

\section{Methods}

Figure~\ref{fig:geom} shows the geometry of a hetero-QDM,
consisting a pair of 3 nm tall InAs dots in the shape of 
truncated cones, grown on two-dimensional InAs wetting layers, embedded in 
a GaAs matrix. The inter-dot separation $d$ is defined as
the distance between
the wetting layers of top and bottom dots. 
We choose the base diameter of top dots (labeled as $\alpha$) to be 20 nm,
and that of the the bottom dots (labeled as $\beta$) to be 19 nm, 
mimicking to the fact that experimentally the top dots are 
slightly larger than the bottom dots. \cite{xie95,solomon96,bodefeld99}
The composition of the dots vary from In$_{0.5}$Ga$_{0.5}$As at their bases
to pure InAs at their top, as determined in Ref. \onlinecite{bayer01}.
We denote the dot molecules made of dissimilar dots
$\alpha$ and $\beta$ as $M_{\alpha\beta}$.  
We also constructed the homo-QDM, consisting a pair of quantum dots
$\gamma$,  which have the average sizes, and the same alloy compositions of 
dots $\alpha$ and $\beta$ in the heteropolar dot molecule. 
We denote the homo-QDM as $M_{\gamma\gamma}$.

The single-particle energy levels and wavefunctions of
$M_{\alpha\beta}$ and $M_{\gamma\gamma}$
are obtained by solving the
Schr\"{o}dinger equations in a pseudopotential scheme,
\begin{equation}
\left[ -{1 \over 2} \nabla^2 
+ V_{\rm ps}({\bf r}) \right] \psi_i({\bf r})
=\epsilon_i \;\psi_i({\bf r}) \; ,
\label{eq:schrodinger}
\end{equation}
where the total electron-ion potential $ V_{\rm ps}({\bf r})$ 
is a superposition of
local, screened atomic pseudopotentials $v_{\alpha}({\bf r})$, and a nonlocal
spin-orbit potential $V_{\rm so}$ i.e.,
$V_{\rm ps}({\bf r}) =\sum_{n,\alpha} 
v_{\alpha}({\bf r} - {\bf R}_{n,\alpha})+V_{\rm so}$.
The atomic position $\{{\bf R}_{n,\alpha}\}$ is obtained from
minimizing the total bond-bending and
bond-stretching energy 
using the Valence Force Field (VFF)
model.~\cite{keating66,martins84}
The atomistic pseudopotentials $v_{\alpha}$ ($\alpha$=In, Ga, As)
are fitted to the physically important quantities of bulk InAs and GaAs, 
including band energies, band-offsets,
effective masses, deformation potentials and alloy
bowing parameters, etc.~\cite{williamson00}
Because for electrons the spin-orbit coupling is extremely small 
in the InAs/GaAs quantum dots, we ignored this effect.
In general, including the spin-orbit coupling effect will introduce
mixture of different total spin states.
Equation (\ref{eq:schrodinger}) is solved in the basis
of $\{\phi_{m,\tensor{\epsilon},\lambda}({\bf k})\}$ of Bloch orbitals of 
band index $m$
and wave vector ${\bf k}$ of material $\lambda$ (= InAs, GaAs), strained
uniformly to strain $\tensor{\epsilon}$ following Ref.~\onlinecite{wang99b}.

The Hamiltonian of interacting electrons
can be written as,
\begin{equation}
H=\sum_{i\sigma} \epsilon_{\alpha}\hat{\psi}^{\dag}_{i\sigma} 
\hat{\psi}_{i\sigma} 
+ {1\over2} \sum_{ijkl} \sum_{\sigma,\sigma'} 
\Gamma^{i,j}_{k,l} 
\hat{\psi}^{\dag}_{i\sigma} \hat{\psi}^{\dag}_{j\sigma'}
\hat{\psi}_{k\sigma'}\hat{\psi}_{l\sigma}\, ,
\label{eq:ham}
\end{equation}
where $\hat{\psi}_{i\sigma}({\bf r}) 
=c_{i\sigma}\psi_{i\sigma}({\bf r})$ is the field operator, whereas 
$c_{i\sigma}$ is a fermion operator. 
$\psi_{i}$= $\sigma_u$, $\sigma_g$, 
$\pi_u$, $\pi_g$ 
are the single-particle eigenfunctions of the 
{\it $i$-th molecular orbital},
and $\sigma$, $\sigma'$=1, 2 are spin indices.
The $\Gamma^{ij}_{kl}$ are 
the Coulomb integrals between molecular orbitals $\psi_{i}$, 
$\psi_{j}$,
$\psi_{k}$ and $\psi_{l}$,
\begin{equation}
\Gamma^{ij}_{kl} =\int\int d{\bf r}d{\bf r'}\; 
{\psi^*_{i}({\bf r}) \psi^*_{j} ({\bf r'}) 
\psi_{k}({\bf r'}) \psi_{l} ({\bf r}) 
\over \epsilon({\bf r}- \bf{r'}) |{\bf r} -{\bf r'}|} \, .
\label{eq:int_m}
\end{equation}
The $J_{ij}=\Gamma^{ij}_{ji}$ and
$K_{ij}=\Gamma^{ij}_{ij}$
are diagonal Coulomb and exchange integrals respectively.
The remaining terms are called off-diagonal or scattering terms.
All Coulomb integrals are calculated numerically from atomistic 
wavefunctions.~\cite{franceschetti99}
We use a phenomenological, position-dependent dielectric  
function $\epsilon({\bf r}- \bf{r'})$ to screen the electron-electron
interaction.\cite{franceschetti99}
The many-particle problem of Eq.(\ref{eq:ham}) is solved via the
CI method, by expanding the $N$-electron wavefunction
in a set of Slater determinants,
$|\Phi_{e_1,e_2,\cdots,e_N}\rangle
=c^{\dag}_{e_1}c^{\dag}_{e_2}\cdots c^{\dag}_{e_N}|\Phi_0\rangle$,
where $c^{\dag}_{e_i}$
creates an electron in the state $e_i$ .
The $\nu$-th many-particle wavefunction is then the linear combinations of
the determinants, 
\begin{equation}
|\Psi_{\nu}\rangle=\sum_{e_1,e_2,\cdots,e_N}
A_{\nu}(e_1,e_2,\cdots,e_N)\;|\Phi_{e_1,e_2,\cdots,e_N}\rangle \; .
\label{eq:coeff}
\end{equation}
For the two-electron problems,
our calculations include all possible Slater 
determinants of six confined molecular orbitals.

\begin{figure}
\includegraphics[width=2.5in,angle=-90]{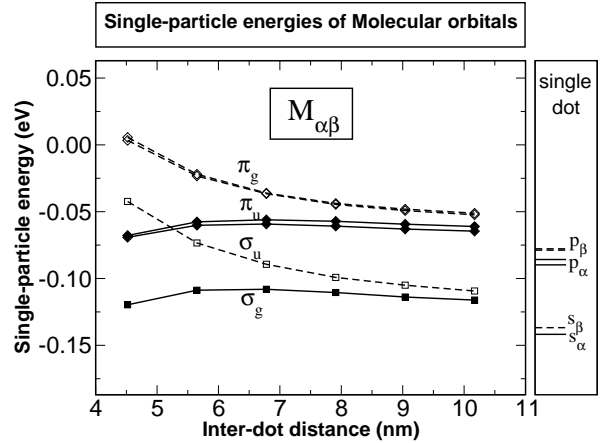}
\caption{Left panel:
The single-particle energy levels of molecular orbitals  {\it vs.} inter-dot distance.
Right panel: The electron single-particle energy levels of the isolated dots 
$\alpha$ and $\beta$.  
}
\label{fig:sp-e}
\end{figure}

\section{Basic electronic structures
    at the single-particle level}

\subsection{Double-dot molecular orbitals}

We first show the electronic structure 
of isolated dots  $\alpha$ and $\beta$. 
The single-dot electron $s$ and $p$
levels of dots $\alpha$ and $\beta$ are shown on the right
panel of Fig.~\ref{fig:sp-e}.  We see that the $s$-$p$ energy spacing of
dot $\alpha$ is 
$\epsilon(p_{\alpha})-\epsilon(s_{\alpha})=$52 meV and that of dot $\beta$ is
$\epsilon(p_{\beta})-\epsilon(s_{\beta})=$59 meV, compared to 54 meV of dot
$\gamma$ (not shown).
The energy level of $s_{\beta}$, 
is slightly ($\sim$ 6 meV) higher
than $s_{\alpha}$, because 
dot $\beta$ is smaller than dot $\alpha$ and therefore has 
larger confinement.
The $p$ levels of all dots have a small energy splitting
due to the underlying
atomistic symmetry, e.g., $\delta\epsilon(p_{\alpha})=$6 meV,
and $\delta\epsilon(p_{\beta})=$1 meV.
We further calculated the fundamental exciton
energy of dot $\alpha$, $E_X(\alpha)=1153$ meV,
and that of dot $\beta$, $E_X(\beta)=1159$ meV.
The energy difference in exciton of dots $\alpha$ and $\beta$
is about 6 meV, in agreement with experiment. \cite{bodefeld99}.
The fundamental exciton energy of the ``averaged''
dot $\gamma$ is $E_X(\gamma) = 1156$ meV.

When two dots $\alpha$ and $\beta$ couple,
the bonding and anti-bonding ``molecular orbitals'' ensue from 
the single-dot orbitals.  
The energy levels of molecular orbitals
are shown on the left panel of Fig. \ref{fig:sp-e}.
We show the single-particle levels of
molecular orbitals\cite{he05b,he05c}
$\sigma_g$, $\sigma_u$ originating from $s$ orbitals, and
$\pi_u$, and $\pi_g$ originating from $p$ orbitals.
The bonding and anti-bonding splitting 
$\Delta_{\sigma}=\epsilon(\sigma_u)-\epsilon(\sigma_g)$ and 
$\Delta_{\pi}=\epsilon(\pi_g)-\epsilon(\pi_u)$ increase with the decrease of
inter-dot distance, 
because the coupling between the top and bottom dots gets stronger.
This picture is similar to what we obtained for 
homo-QDMs.   
However, there is an important difference between the homo-QDMs $M_{\gamma\gamma}$ 
and hetero-QDMs $M_{\alpha\beta}$ : 
in the former case, the bonding and anti-bonding splitting $\Delta_{\sigma}$
and  $\Delta_{\pi}$ decay to almost zero at
large inter-dot distance, 
while in the later case, $\Delta_{\sigma}$
and  $\Delta_{\pi}$ tend to constants ($\Delta_{\sigma}\sim$ 7 meV, 
$\Delta_{\pi}\sim$ 10 meV here),
because the molecular orbitals 
gradually convert at large inter-dot distance to
single dot energy levels, e.g. the $\sigma_g$
levels convert to top dot $s$ orbitals, and 
$\sigma_u$ convert to bottom dot $s$ orbitals, therefore
the energy splitting between the first and second molecular states at large distances 
is approximately the energy difference between $s$ orbitals of the top and bottom dots, i.e.,   
$\Delta_{\sigma}\sim \epsilon(s_{\beta})-  \epsilon(s_{\alpha})\neq 0$ 
for $M_{\alpha\beta}$.

Figure \ref{fig:sp-e} shows that at inter-dot distance $d=$10 nm, 
the molecular orbital levels are 
about 25 meV higher than the isolated dot levels, although the direct
electronic coupling between two dots is much smaller than this quantity. 
This energy shift results from 
the long range strain effects experienced by one dot
due to the presence of the second dots. This effect is missed in EMA-type
model calculations, \cite{rontani01} which ignore strain effects.

\subsection{Single dot-localized orbitals}
\label{sec:local-orbitals}

The above discussions pertain to the basis of double-dot molecular orbitals. 
An alternative way to study QDMs is to use a dot-localized basis.
We have demonstrated \cite{he05b,he05c} that dot-localized orbitals 
can be a useful tool to analyze the QDM physics, 
including the electron double
occupation, and two-electron entanglement.

Dot-localized orbitals $\chi_{\eta}$
can be obtained from a unitary rotation of molecular orbitals, i.e.,
\begin{equation}
\chi_{\eta} =\sum^N_{i=1} \mathcal{U}_{\eta\, ,i}\; \psi_{i} \, ,
\end{equation}
where, $\psi_{i}$ is the $i$-th molecular orbital, and 
$\mathcal{U}$ is a unitary matrix, i.e., $\mathcal{U}^{\dag}\mathcal{U}=I$.
We choose the unitary matrices $\mathcal{U}$ that maximize the total orbital
self-Coulomb energy.~\cite{edmiston63,he05c} 
The procedure of finding $\mathcal{U}$ is 
described in the Appendix B of Ref.\onlinecite{he05c}.
As we will show below 
these dot-localized orbitals $\chi_{\eta}$ have the advantage of 
being only weakly dependant to the inter-dot coupling. 
This invariance may provide
simplified pictures for qualitatively understanding of the QDM physics.

\begin{figure}
\includegraphics[width=2.3in,angle=-90]{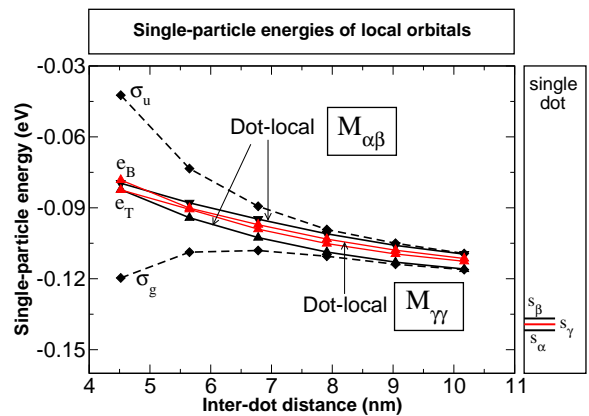}
\caption{ (Color online) Left panel: The energy levels 
of dot-localized orbitals
for QDMs $M_{\alpha\beta}$ (black solid lines)
and $M_{\gamma\gamma}$ (red solid lines).
$e_T$ and $e_B$ denote the $s$ orbitals of the top and bottom dots
respectively. The molecular orbitals energy levels $\sigma_g$
and $\sigma_u$ (dashed lines) are shown for
dot molecules $M_{\alpha\beta}$.  
Right panel: $s$ levels of {\it isolated} dots $\alpha$, $\beta$ and $\gamma$.
}
\label{fig:sp-local}
\end{figure}

\subsubsection{Single-particle energies of dot-localized orbitals}

The single-particle levels of dot-localized orbitals and
the hopping (or tunneling) term between two dots can be obtained from
\begin{eqnarray}
e_{\eta} &=&
\langle \chi_{\eta} | \hat{H}_0|\chi_{\eta} \rangle 
=\sum_{i} \mathcal{U}^*_{\eta\, ,i}\, \mathcal{U}_{\eta\, ,i}\, 
\epsilon_{i} \; ,\\
\label{eq:epsilon}
t_{\eta_1\eta_2} &=& \langle \chi_{\eta_1} | \hat{H}_0|\chi_{\eta_2} \rangle
 =\sum_{i} \mathcal{U}^*_{\eta_1,\, i}\, U_{\eta_2,\, i}\, 
\epsilon_i \; , 
\label{eq:t}
\end{eqnarray}
where, $\epsilon_{i}$ is the single-particle energy of $i$-th molecular
orbital and $\hat{H}_0=\sum_{i\sigma} \epsilon_{\alpha}\hat{\psi}^{\dag}_{i\sigma} 
\hat{\psi}_{i\sigma}$ is the single-particle Hamiltonian.
Figure~\ref{fig:sp-local} depicts
the single-particle levels $e_T$ and $e_B$ of the dot-localized
orbitals of both top and bottom dots,
for inter-dot distances $d$ in the range from 4 nm to 10 nm.
(Here, we denote the top dot $T$ and the bottom dot $B$, 
to distinguish them from isolated dots $\alpha$, $\beta$ and $\gamma$).
$e_T$ and $e_B$ of $M_{\alpha\beta}$ are shown in the black solid lines,
and those of $M_{\gamma\gamma}$ are shown in the red solid lines,
At large $d$, the energy difference $e_B-e_T \sim 6$ meV
for $M_{\alpha\beta}$, is close to the value of
difference $\epsilon(s_{\beta})-\epsilon(s_{\alpha})$
between $s$ orbitals of isolated dots $\alpha$, $\beta$.
This energy difference gets smaller when the two dots move closer, 
because the energy levels of the top dot rise faster than those of bottom
dots due to the strain asymmetry. 
For the homo-QDMs $M_{\gamma\gamma}$, $e_T$ and $e_B$ are almost
degenerate. The small difference  ($\sim$ 1 meV) between them is due to the
strain and alloy effects.
%
We also plot in Fig.~\ref{fig:sp-local} the energies of molecular orbitals
$\sigma_u$ and $\sigma_g$ in dashed lines for $M_{\alpha\beta}$. As we see,
for $d>$9 nm, the dot-localized state $e_B$ of $M_{\alpha\beta}$
is almost identical to the molecular orbital  
$\sigma_u$, while $e_T$ merges
with $\sigma_g$, indicating at large $d$, molecular orbitals convert to
dot-centered orbitals for $M_{\alpha\beta}$.

\begin{figure}
\includegraphics[width=2.5in,angle=-90]{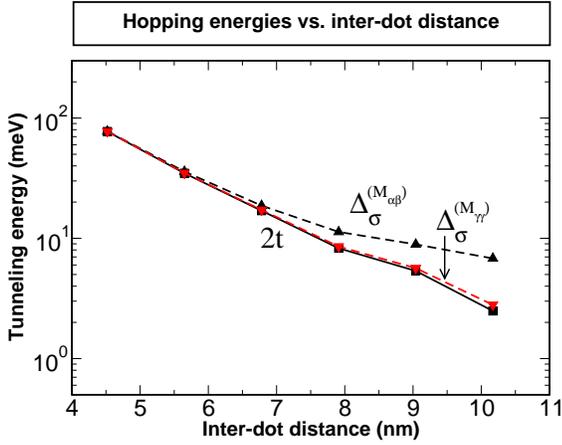}
\caption{ (Color online) The inter-dot hopping energy 2$t$ (solid lines)
of hetero-QDM $M_{\alpha\beta}$ 
and homo-QDM $M_{\gamma\gamma}$.
We also show the bonding-antibonding 
splitting $\Delta_{\sigma}$ of $M_{\alpha\beta}$ and $M_{\gamma\gamma}$ .
}
\label{fig:tunneling}
\end{figure}

The quantity 2$t$ measures the coupling strength 
between the top and bottom dots,
and directly determines the two-electron properties such as singlet-triplet
splitting in the QDM. 
We calculate this hopping energy between the $s$ orbitals of
top and bottom dots at different inter-dot distances 
for both $M_{\alpha\beta}$ and $M_{\gamma\gamma}$
in Fig.~\ref{fig:tunneling}. 
(We ignore the orbital index ``$s$'' to simplify the notation.)
We find that 2$t(M_{\alpha\beta})$ and
2$t(M_{\gamma\gamma})$ are almost 
identical at all inter-dot distance. 
However, the hopping energies calculated here are much larger than 
we obtained for the pure InAs/GaAs QDM \cite{he05c}, 
because the alloy QDM have much smaller energy barrier between two dots 
than pure QDM.
In general, the quantity 2$t$ does not equal to 
the bonding-antibonding splitting
$\Delta_{\sigma} = \sqrt{\delta^2+4t^2}$, where 
$\delta=\epsilon(e_T)-\epsilon(e_B)$, being the energy difference
of $s$ orbitals of the top and bottom dots.
For homo-QDMs, where $\delta/2t \ll 1$, we have 2$t\sim \Delta_{\sigma}$
as seen in Fig. \ref{fig:tunneling}. 
However, for hetero-QDMs, $\Delta_{\sigma}$ may be significantly
different from 2$t$, especially at {\it large} inter-dot distances, 
where $\delta/2t \gg 1$,
also illustrated in Fig.~\ref{fig:tunneling}. 
Experimentally,\cite{ota05} one usually measures 
the bonding-antibonding splitting
rather than the hopping 2$t$. Therefore, to get the hopping energy
between two dots, one need to know the energy difference $\delta$
of two dots.

\begin{figure}
\includegraphics[width=2.6in,angle=-90]{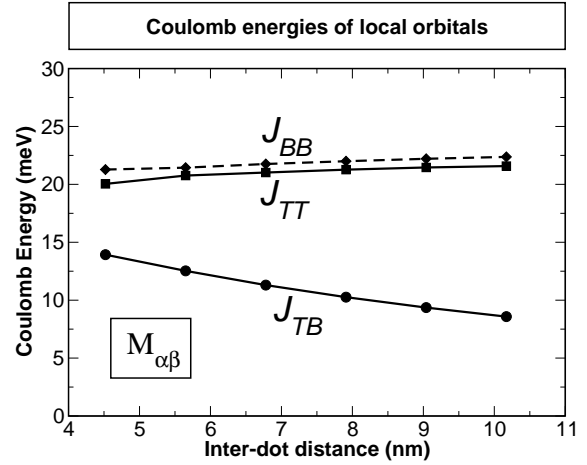}
\caption{The Coulomb energies of dot-localized orbitals
of hetero-QDM $M_{\alpha\beta}$. $J_{TT}$ and
  $J_{BB}$ are the $s$ orbital self-Coulomb energies of top and bottom dots
  respectively, whereas $J_{TB}$ are the Coulomb energies between $s$
  orbitals of the top and the bottom dots.
}
\label{fig:JK-local}
\end{figure}

\subsubsection{Coulomb integrals of dot-localized orbitals}

The Coulomb integrals in the dot-localized basis can be obtained from 
Coulomb integrals of molecular orbitals as follows,
\begin{equation}
\widetilde{\Gamma}^{\eta_1,\eta_2}_{\eta_3,\eta_4} 
=\sum_{i,j,k,l}
\mathcal{U}^*_{\eta_1\, ,i}\, \mathcal{U}^*_{\eta_2\,,j}\,
\mathcal{U}_{\eta_3\, ,k}\,\mathcal{U}_{\eta_4\, ,l}\; 
\Gamma^{i,j}_{k, l} \, ,
\label{eq:col-int}
\end{equation} 
where $\Gamma^{i,j}_{k, l}$ are the Coulomb integrals in the molecular basis.
The direct Coulomb integrals $J_{TT}$, $J_{BB}$ and $J_{TB}$ for 
$M_{\alpha\beta}$
are shown in Fig.~\ref{fig:JK-local}.
The Coulomb integrals $J_{TT}\sim J_{\alpha\alpha}$=21.4 meV 
and $J_{BB}\sim J_{\beta\beta}$=22.3 meV, 
are almost constants at all inter-dot distances, suggesting that 
the dot-localized orbitals are approximately unchanged for different inter-dot
distance $d$. $J_{\beta\beta} > J_{\alpha\alpha}$, as
dot $\beta$ is smaller than dot $\alpha$. 
The inter-dot Coulomb interaction $J_{TB}$ decay slowly as $1/d$.
The exchange energies (not shown) between the top and bottom electrons is 
orders of magnitude smaller than the hopping energy, 
and therefore can be ignored in practice. 
For the homo-QDM $M_{\gamma\gamma}$, we found that on-site Coulomb energies
$J_{TT}$ $\sim$ $J_{BB}$,
both are very close to the average values of $J_{TT}$ and $J_{BB}$ 
of $M_{\alpha\beta}$. 
The inter-dot Coulomb energies $J_{TB}$ of
$M_{\alpha\beta}$ and  $M_{\gamma\gamma}$ are also extremely close. 

\begin{figure}
\includegraphics[width=3.0in,angle=0]{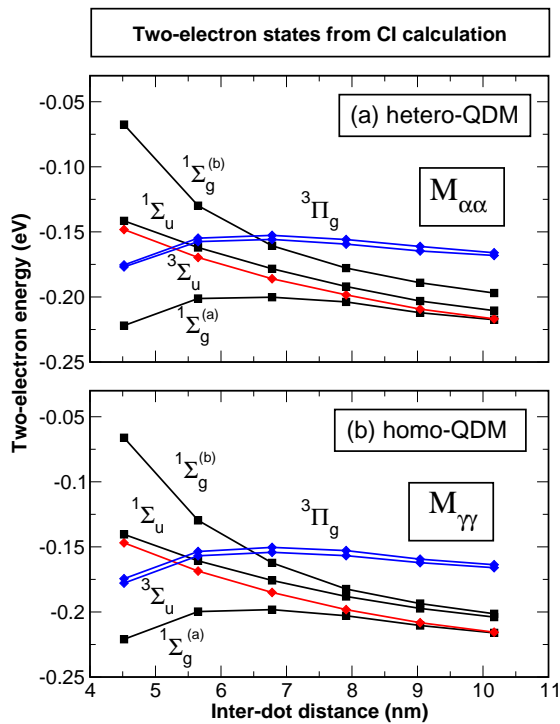}
\caption{(Color online) Two-electron states for (a) hetero-QDM $M_{\alpha\beta}$ 
and (b) homo-QDM $M_{\gamma\gamma}$,
including the singlet $^1\Sigma_g^{(a)}$, $^1\Sigma_u$,
$^1\Sigma_g^{(b)}$ states and the 3-fold degenerated
triplet states $^3\Sigma_u$ as well as two 3-fold degenerated
triplet states $^3\Pi_u$.
}
\label{fig:CI}
\end{figure}

\section{Two electrons in the dot molecule}

\subsection{Many-body energy states}

The two-electron-in-a-QDM problem is of special interest, as
 it is the prototype of quantum gate using QDMs.\cite{loss98}
We calculate the two-electron energy levels 
by the configuration interaction 
method using Slater determinants constructed from 
confined molecular orbitals $\sigma_g$, $\sigma_u$ and
$\pi_u$, $\pi_g$, which give 66 configurations in total. 
The two-electron energies $\Sigma$ and $^3\Pi_u$
for hetero-QDMs $M_{\alpha\beta}$ 
are plotted in Fig.~\ref{fig:CI}(a).
To compare with homo-QDMs, 
we show the two-electron states of $M_{\gamma\gamma}$
in  Fig.~\ref{fig:CI}(b).
The energy levels of $M_{\alpha\beta}$  are similar to those
of $M_{\gamma\gamma}$, in the following way:
(i) The order of the CI levels is unchanged, particularly 
the ground states are still the singlet states $^1\Sigma^{(a)}_g$ at all
inter-dot distance; 
(ii) The trend of each CI level vs. inter-dot distance $d$
is similar to what we obtained for $M_{\gamma\gamma}$.
There are also some differences between the hetero-QDMs
$M_{\alpha\beta}$ and homo-QDMs $M_{\gamma\gamma}$, especially at 
larger inter-dot distances.
For example, in the homopolar QDMs, the $^1\Sigma_u$ state is
almost degenerate with $^1\Sigma_g^{(b)}$ at large inter-dot distance, while
in $M_{\alpha\beta}$, $^1\Sigma_g^{(b)}$  is about 13 meV higher than 
$^1\Sigma_u$ at $d$=10 nm. At large $d$, 
$^1\Sigma_u$ and $^1\Sigma_g^{(b)}$ 
correspond to the states that 
two electrons localized on the same dots. \cite{he05b, he05c} 
The energy difference between  $^1\Sigma_g^{(b)}$ and $^1\Sigma_u$ 
is due to the size difference of dots $\alpha$ and $\beta$.

\begin{figure}
\includegraphics[width=2.8in,angle=-90]{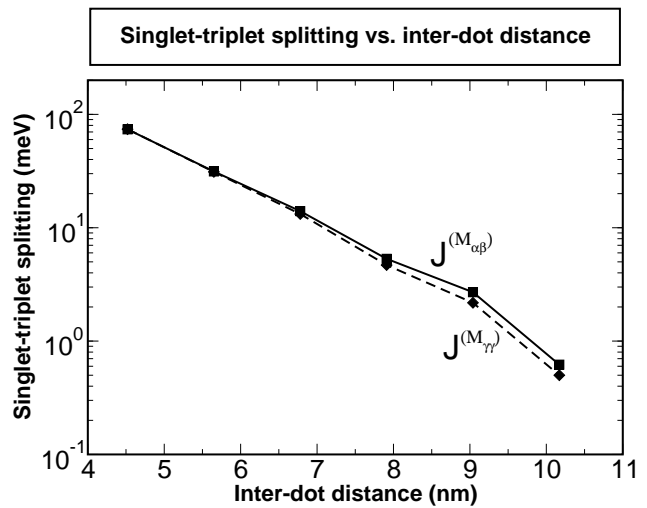}
\caption{The singlet-triplet splitting $J_{S-T}$ vs
  {\it vs.} inter-dot distance for hetero-QDM
$M_{\alpha\beta}$ (solid line) and homo-QDM
$M_{\gamma\gamma}$ (dashed line).
}
\label{fig:Jst}
\end{figure}

The singlet $^1\Sigma_g^{(a)}$ 
and triplet states $^3\Sigma$ can be used as two qubit states in quantum
computing. 
In a proposed quantum SWAP gate,\cite{loss98} the gate operation time
$\tau \sim 1/J_{S-T}$, where $J_{S-T}$
being the singlet-triplet energy splitting. 
The singlet-triplet splitting of $M_{\alpha\beta}$
is shown in Fig.~\ref{fig:Jst} on a semi-log
plot. We see that it decay approximately exponentially with the inter-dot
distance.
We also show in Fig.~\ref{fig:Jst}
the singlet-triplet splitting of the homo-QDM $M_{\gamma\gamma}$. 
We found that the $J_{S-T}$ of homo-QDM $M_{\gamma\gamma}$ is slightly smaller 
than the $J_{S-T}$ of hetero-QDM $M_{\alpha\beta}$, though the hopping
energies of $M_{\alpha\beta}$ and $M_{\gamma\gamma}$ are almost identical.
In the hetero-QDM case, the singlet wavefunction has more weight
on the lower energy dot and therefore lowers the singlet energy
and increases the singlet-triplet splitting.

\subsection{Double occupation of one of the dots in a QDM}
\label{sec:docc}

Double occupation means that two electrons occupy the same dot in a QDM. 
If the double occupation rate is high, the quantum gate operation may fail.
The double occupation rate also reflects the
localization properties of electrons in the QDM. 
If the double occupation rate is zero, each dot has one electron, 
whereas double occupation rate of 1 means
that two electrons are always localize on a single dot. 
When the double occupation rate is 0.5, two electrons are delocalized
between two dots.
The double occupation can be conveniently analyzed in the 
dot-localized basis by transforming
the CI equations to the dot-localized basis. \cite{he05b}
In the simplest case, we consider only the ``s'' orbital for each dot,  
which give six configurations as follows,
 $|e_T^{\uparrow}, e_B^{\uparrow} \rangle$,
$|e_T^{\downarrow}, e_B^{\downarrow} \rangle$,
$|e_T^{\uparrow}, e_B^{\downarrow} \rangle$,  
$ |e_T^{\downarrow}, e_B^{\uparrow} \rangle$,
$|e_B^{\uparrow}, e_B^{\downarrow} \rangle$
and 
$|e_T^{\uparrow}, e_T^{\downarrow} \rangle$.
The Hamiltonian in this basis set is, \cite{he05c}
\begin{widetext}
\begin{equation}
\scriptsize
H= \left(\begin{array}{cccccc}
e_{\rm T}+e_{\rm B}+J_{\rm TB}-K_{\rm TB} 
&  0 &   0&   0  &0  & 0\\
0  & e_{\rm T}+e_{\rm B}+J_{\rm TB}
-K_{\rm TB} &  0&  0  &0 & 0 \\
0  &  0 &e_{\rm T}+e_{\rm B}
+J_{\rm TB} &-K_{\rm TB}&t
& t \\
0  & 0   &-K_{\rm TB} &  e_{\rm T}
+e_{\rm B}+J_{\rm TB} &-t
& -t \\
0  & 0   &  t
& -t
& 2e_{\rm B}+J_{\rm BB}  &0\\
0  & 0   &  t
& -t  & 0    
& 2e_{\rm T} +J_{\rm TT}
\end{array}\right)\;.
\normalsize
\label{eq:hubbard}
\end{equation}
\end{widetext}
where $t=t_{TB}$.
We ignored in Eq.(\ref{eq:hubbard}) the
off-diagonal Coulomb integrals, which are much smaller than the hopping $t$.

\begin{figure}
\includegraphics[width=3.0in,angle=0]{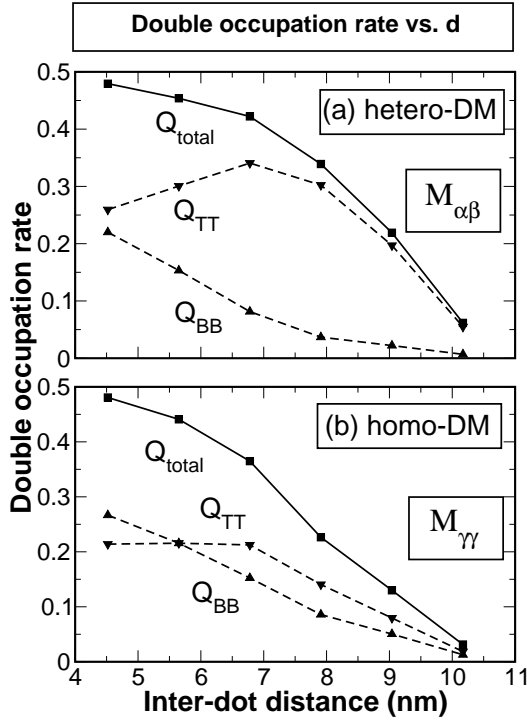}
\caption{The double occupation rate of the ground-state singlet
 $^1\Sigma_g^{(a)}$ 
{\it vs.} inter-dot distance for (a) hetero-QDM $M_{\alpha\beta}$ 
and (b) homo-QDM $M_{\gamma\gamma}$.
}
\label{fig:d_occ_s1}
\end{figure}

The calculation of the matrix elements of
Eq. (\ref{eq:hubbard}) is described in Sec.\ref{sec:local-orbitals}.
The two electrons can be either both  on the top dots, or both on the bottom
dots, or one on the top and the other on the bottom dots. 
We denote by $\{|\chi_{l,p}^{\sigma},\chi_{l',p'}^{\sigma'}\rangle\}$
the configuration where one electron is on the $l$-th orbital of the $p$ dot 
with spin $\sigma$, and the other electron is on the 
$l'$-th orbital of the $p'$ dot 
with spin $\sigma'$.
Then the double occupation rate 
$Q^{(\nu)}_{pp}$ in the
many-particle state $\nu$ is the probability of 
two electrons occupying the dot $p$= (T or B) at the same time, i.e.,
\begin{equation}
Q^{(\nu)}_{pp}=\sum_{l\sigma,l'\sigma'} P_{\nu}(|\chi^{\sigma}_{l,p},
\chi^{\sigma'}_{l',p}\rangle) \, ,
\label{eq:d_occ}
\end{equation}
where $P_{\nu}(\mathcal{C})$ is the weight of the configuration $\mathcal{C}$ 
in the many-body wave functions of state $\nu$. The total probability
of two electrons being on the {\it same} dot is then 
$Q^{(\nu)}_{\rm tot}
=Q^{(\nu)}_{\rm TT}+Q^{(\nu)}_{\rm BB}$ for the $\nu$-th
state.   

We plot $Q_{\rm tot}$, $Q_{\rm TT}$ and $Q_{\rm BB}$ of state
$^1\Sigma_g^{(a)}$ for $M_{\alpha\beta}$ in Fig. \ref{fig:d_occ_s1}(a)
and for $M_{\gamma\gamma}$ in Fig. \ref{fig:d_occ_s1}(b).
We also performed calculations on a ``symmetrized'' model QDM
$M_{\alpha'\alpha'}$ by setting $e'_T=e'_B=(e_T+e_B)/2$
and $J'_{TT}=J'_{BB}=(J_{TT}+J_{BB})/2$ of $M_{\alpha\beta}$ 
in Eq. (\ref{eq:hubbard}). 
$M_{\alpha'\alpha'}$ represents an {\it ideal} homo-QDM,
without the asymmetry caused by strain, size and alloy composition effects. 
When compare the double occupation of the hetero- and homo-QDMs, 
we see that
(i) For both types of QDMs, 
$Q_{\rm tot} \sim $ 0.5 at $d \sim$ 4.5 nm,
meaning that two electrons are delocalized on two dots. 
For both QDMs, $Q_{\rm tot}$ 
decays monotonically with the inter-dot distance, and at 
$d \sim $ 10 nm, $Q_{\rm tot} \sim $ 0,  meaning that the two
electrons are about each localized on one of the two dots.\\
On the other hand, the double occupation of individual 
dot $Q_{\rm TT}$ and $Q_{\rm BB}$ differ
substantially for homo-QDMs and hetero-QDMs: \\ 
(ii) For the homo-QDM $M_{\alpha'\alpha'}$, 
$Q_{\rm BB}= Q_{\rm TT}$ and
decay monotonically with the inter-dot distances.
$Q_{\rm BB}$ and $Q_{\rm TT}$ of $M_{\gamma\gamma}$ have similar features,
although $Q_{\rm BB}$ is slightly different from $Q_{\rm TT}$ due to the
strain and alloy effects. This feature is also seen in the homo-QDM made of pure
InAs/GaAs dots \cite{he05b, he05c}.
In the hetero-QDMs $M_{\alpha\beta}$, 
$Q_{\rm TT}$ behaves very differently from $Q_{\rm BB}$ because the effective
single-particle energy $e_T < e_B$. 
Whereas $Q_{\rm BB}$ decays monotonically with the inter-dot distance,
$Q_{\rm TT}$ has a maximum at $d \sim$ 7 nm. 
The reason is that at $d \sim$ 4.5 nm, 
the hopping energy 2$t$ is much larger than
$e_B-e_T$, therefore the electrons can overcome the energy barrier
between the top and bottom dots and distribute evenly between two dots,
leading to $Q_{\rm TT} \sim Q_{\rm BB}$. 
At larger $d$, $2t \ll e_B-e_T$, and the electrons would prefer to 
localize on the top dots, 
leading to $Q_{\rm TT} \gg Q_{\rm BB}$. Therefore, even when
the total double occupation rate drops down, $Q_{\rm TT}$ still increases and
reaches the maximum at $d$=7 nm. For $d >$7 nm, $Q_{\rm TT}$ decays 
as $Q_{\rm tot}$ decays. \\
(iii) The homo-QDMs $M_{\alpha'\alpha'}$ and $M_{\gamma\gamma}$ have almost
the same {\it total} double occupation, both smaller than that of the hetero-QDM
$M_{\alpha\beta}$.
The asymmetry between two dots increases the total double
occupation. In an extreme case, where $e_T \ll e_B$, the two electrons could
always localize on the top dots, leading to $Q_{\rm tot}$=$Q_{\rm TT}$=1.

\begin{figure}
\includegraphics[width=2.8in,angle=0]{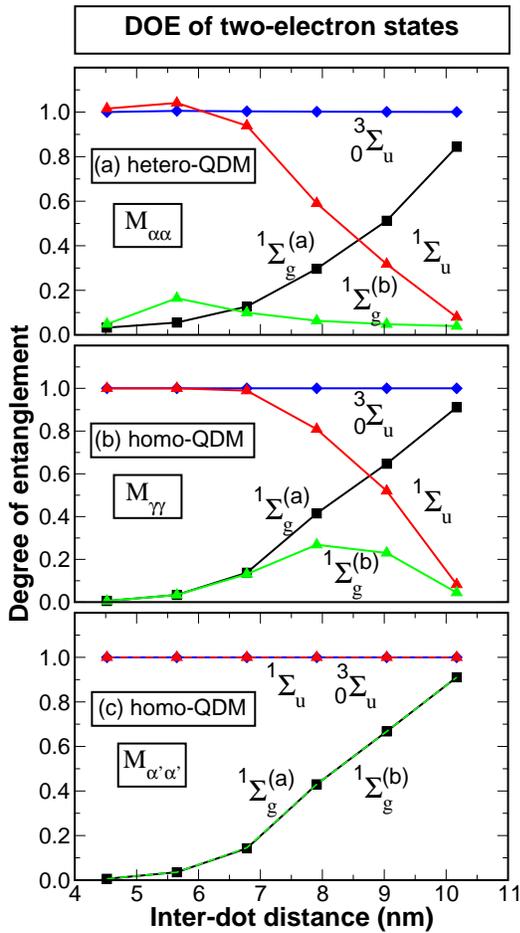}
\caption{(Color online) The degree of entanglement of two-electron states
$^1\Sigma_g^{(a)}$, $^1\Sigma_u$,  $^1\Sigma_g^{(b)}$  
and $^3\Sigma_u$, in (a) the hetero-QDM $M_{\alpha\beta}$,
(b) the homo-QDM $M_{\gamma\gamma}$ and 
(c) the model ``symmetrized'' homo-QDM $M_{\alpha'\alpha'}$.
}
\label{fig:DOE}
\end{figure}

\section{Entanglement}
\label{sec:doe}

\subsection{Degree of entanglement for two electrons}

The degree of entanglement (DOE) is one of the most 
important quantities for
successful quantum gate operations. 
For {\it distinguishable} particles such as 
an electron and a hole, the DOE can be
calculated from the Von Neumann-entropy formulation.
\cite{nielsen_book,bennett96,bennett96b,wehrl78} 
However, Von Neumann entropy formulation can not be used directly to calculate
DOE for {\it indistinguishable} particles.
\cite{schliemann01a,paskauskas01,li01,zanardi02,
shi03,wiseman03,ghirardi04}
Schliemann et al. proposed a quantum correlation function for two electrons 
which has similar properties as the DOE. \cite{schliemann01a} However, the
generalization of this quantum correlation function to a system that has more
than two single-particle levels is complicated. 
We proposed a DOE measure \cite{he05c} for 
indistinguishable fermions using the Slater decompositions 
\cite{yang62,schliemann01a} as, 
\begin{equation}
\mathcal{S}=-\sum_{u} z_{i}^2 \;{\rm log}_2\, z_{i}^2\; ,
\label{eq:my-entropy-t}
\end{equation}
where, $z_i$ are Slater decomposition coefficients
and $\sum_i z_i^2$=1.
As shown in Ref. \onlinecite{he05c}, 
the DOE measure Eq.(\ref{eq:my-entropy-t}) reduces to the usual 
Von Neumann entropy for {\it distinguishable} 
particles when the two electrons
are far from each other. In Refs. \onlinecite{paskauskas01,li01},
a similar DOE measure was defined, which however
due to a different normalization
condition for $z_i$ was used, does not reduce to 
the usual Von Neumann entropy even when the two electrons can be distinguished
by their sites.

The DOE of $\Sigma$ states calculated from Eq. (\ref{eq:my-entropy-t}) for 
the hetero-QDM $M_{\alpha\beta}$, the homo-QDM  $M_{\gamma\gamma}$,
and the model homo-QDM $M_{\alpha'\alpha'}$
are shown in Fig. \ref{fig:DOE}(a), (b) and (c) respectively.
All of the three QDMs have the following features:
(i) $\mathcal{S}(^1\Sigma_g^{(a)})$ is close to zero (unentangled)
at $d\sim$ 4.5 nm, and close to unity (fully entangled)
at $d \sim$ 10 nm.
(ii) $\mathcal{S}(^3_0\Sigma)$ is almost unity (fully entangled)
at all inter-dot distances.
However, $\mathcal{S}(^1\Sigma_g^{(a)})$ of the homo-QDM $M_{\gamma\gamma}$ 
(which is very close the $\mathcal{S}(^1\Sigma_g^{(a)})$ of
$M_{\alpha'\alpha'}$) is larger than 
$\mathcal{S}(^1\Sigma_g^{(a)})$ of the hetero-QDM $M_{\alpha\beta}$,
showing that the asymmetry in a QDM 
lowers the two-electron entanglement of the ground state singlet.  
%
%

In contrast to $\mathcal{S}(^1\Sigma_g^{(a)})$ and 
$\mathcal{S}(^3_0\Sigma)$,  
$\mathcal{S}(^1\Sigma_g^{(b)})$ and $\mathcal{S}(^1\Sigma_u)$
are very sensitive to the asymmetry of the QDMs. 
In general, if the
two dots have identical electronic structures
(e.g., in the simple Hubbard model), 
$\mathcal{S}(^1\Sigma_g^{(b)})=\mathcal{S}(^1\Sigma_g^{(a)})$
and $\mathcal{S}(^1\Sigma_u)$=1, \cite{he05c} as is illustrated
in Fig. \ref{fig:DOE}(c) for $M_{\alpha'\alpha'}$.
For $M_{\gamma\gamma}$, which is somehow asymmetric due to the strain 
and alloy effects, 
$\mathcal{S}(^1\Sigma_g^{(b)})$ is close to
$\mathcal{S}(^1\Sigma_g^{(a)})$ at small $d$, and drops down at large $d$,
whereas for $M_{\alpha\beta}$,
$\mathcal{S}(^1\Sigma_g^{(b)})$ is different from $\mathcal{S}(^1\Sigma_g^{(a)})$
at all inter-dot distances.
The slight asymmetry in $M_{\gamma\gamma}$ also causes
$\mathcal{S}(^1\Sigma_u)$ to drop down at large $d$, 
similar to $\mathcal{S}(^1\Sigma_u)$ of $M_{\alpha\beta}$.

\begin{figure}
\includegraphics[width=2.5in,angle=-90]{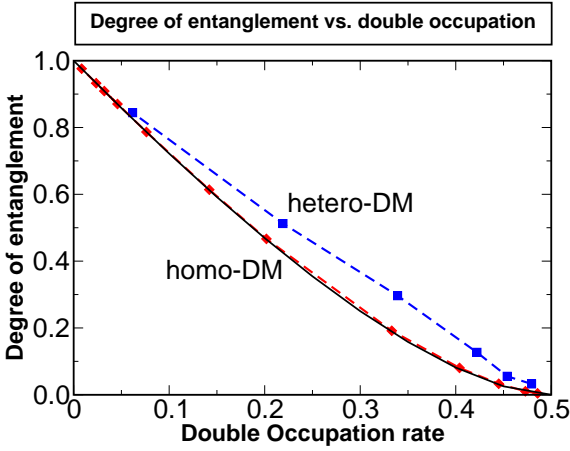}
\caption{(Color online) Comparison of the degree of entanglement {\it vs.}
 double occupation rate for hetero- and homo-QDMs. 
The black solid line represents the
analytical results of homo-QDM, and the red dashed line represents the 
numerical results for
homo-QDM $M_{\alpha'\alpha'}$, and 
$M_{\gamma\gamma}$, whereas the blue line represents the results 
for hetero-QDMs $M_{\alpha\beta}$.
}
\label{fig:occ_DOE}
\end{figure}

\subsection{Degree of entanglement vs double occupation}

Experimentally, it is very hard to measure the DOE of two electrons in the
QDM directly, while it is relatively easy to measure the possibility
of double occupation. 
Therefore it would be useful to explore the relation between DOE and 
the double occupation rate.
The triplet states $^3\Sigma$ have negligible double occupation rate
due to the Pauli exclusion principle. 
%
Here, we discuss the relation between DOE and double occupation rate
for the ground state singlet $^1\Sigma_g^{(a)}$.
We consider the simplest case, where only ``s'' orbital in each dot is
considered.
The ground state singlet $^1\Sigma_g^{(a)}$ wavefunction can be generally
written as,
\begin{equation}
\Psi(^1\Sigma_g^{(a)})=c_1 |e_T^{\uparrow}, e_B^{\downarrow} \rangle
              + c_2 |e_B^{\uparrow}, e_T^{\downarrow} \rangle
              + c_3 |e_T^{\uparrow}, e_T^{\downarrow} \rangle
              + c_4 |e_B^{\uparrow}, e_B^{\downarrow} \rangle \, ,
\label{eq:psi}
\end{equation}
and $|c_1|^2+|c_2|^2+|c_3|^2+|c_4|^2=1$.
Alternatively, we have 
\begin{equation}
\Psi(^1\Sigma_g^{(a)})=\sum_{i,j} \omega_{ij} |i\rangle \otimes |j\rangle \;
\label{eq:wav}
\end{equation}
where,
\begin{equation}
\omega=
\left(\begin{array}{cccc}
0  & -c_3 & 0 & -c_1\\
c_3 & 0 & c_2 & 0 \\
0  & -c_2 & 0 & -c_4\\
c_1 & 0 & c_4 &0
\end{array}\right)\;,
\label{eq:omega-s1-1}
\end{equation}
and $|i\rangle$ ,$|j\rangle$ =  
$|e_T^{\uparrow}\rangle$, $|e_T^{\downarrow}\rangle$, 
$|e_B^{\uparrow}\rangle$, $|e_B^{\downarrow}\rangle$.
We can use Eq. \ref{eq:my-entropy-t} to calculate the DOE, 
where $z_1^2= 1/2(1-\sqrt{1-4(c_1c_2-c_3c_4)^2}\,)$
and $z_2^2= 1/2(1+\sqrt{1-4(c_1c_2-c_3c_4)^2}\,)$ 
are the eigenvalues of $\omega^{\dag}\, \omega$.
For a QDM with reflection symmetry, we have
$c_1=c_2$ and $c_3=c_4$, 
and therefore
$z_1^2=1/2(1-\sqrt{1-(1-4c_3^2)^2}\;)$, and 
$z_2^2=1/2(1+\sqrt{1-(1-4c_3^2)^2}\;)$.
Using the definition of double occupation rate, 
$Q_{\rm tot}=c_3^2+c_4^2$, we have
\begin{eqnarray}
z_1^2 & = & 1/2(1-\sqrt{1-(1-2Q_{\rm tot})^2}\,)\, , \nonumber \\
z_2^2 & = & 1/2(1+\sqrt{1-(1-2Q_{\rm tot})^2}\,)\, .
\label{eq:z1-z2}
\end{eqnarray}
The DOE of $^1\Sigma_g^{(a)}$ is calculated by substituting
$z_1^2$, $z_2^2$ into Eq. (\ref{eq:my-entropy-t}).
We plot the DOE vs. double occupation rate of the above ideal model
in Fig.~\ref{fig:occ_DOE} in a black solid line. We also present in the same
figure, the DOE 
of $M_{\alpha\beta}$, $M_{\gamma\gamma}$ and  $M_{\alpha'\alpha'}$
vs. double occupation rate. 
We found that the double occupation dependence of DOE for
the homo-QDM $M_{\alpha'\alpha'}$ has perfect
agreement with the analytical result, which is also true 
for $M_{\gamma\gamma}$ even though it has small
asymmetry in the molecule due to the strain and alloy effects.
We also checked the homo-QDM made 
of pure InAs/GaAs dots \cite{he05b, he05c},
and found the same double occupation dependence of DOE for the
$^1\Sigma_g^{(a)}$ state, indicating this is a robust feature for homo-QDMs.
However, the double occupation dependence of DOE for
$M_{\alpha\beta}$ deviates from the ideal case because 
dots $\alpha$ and $\beta$ are different.

\section{Summary}

We have studied the electronic structures of quantum dot molecules
made of (In,Ga)As/GaAs dots of different sizes (hetero-QDMs), and compare them to that
of quantum dot molecules made of identical dots (homo-QDMs). 
We found that while
the hetero-QDMs and homo-QDMs have relatively similar electronic structures at
short inter-dot distance, they differ significantly at large inter-dot distance.
(i) Unlike those of homo-QDMs, the  single-particle molecular orbitals 
of hetero-QDMs convert to dot localized orbitals at large inter-dot distance. 
(ii) Consequently, the bonding-antibonding splitting of molecular orbitals is
significantly larger than the electron  hopping energy in a hetero-QMD at
large inter-dot distance, whereas
for homo-QDM, the bonding-antibonding splitting is very similar to the 
hopping energy.
(iii) The asymmetry of the QDM will significantly increase the double
occupation for the two-electron ground states, and therefore lowers the degree of
entanglement of the two electrons.

\acknowledgments
L. He acknowledges the support from the Chinese National
Fundamental Research Program, the Innovation
funds and ``Hundreds of Talents'' program from Chinese Academy of
Sciences, and National Natural Science Foundation of China (Grant
No. 10674124).
The work done at NREL was funded by the U.S. Department of Energy, Office of Science,
Basic Energy Science, Materials Sciences and Engineering, LAB-17 initiative,
under Contract No. DE-AC36-99GO10337 to NREL.


\end{document}